\def\~{{$\tilde{\phantom{a}}$}}

\documentclass [12pt] {article}
\usepackage{epsfig}
\usepackage{color}

\def\thebibliography#1{\section{References}\markboth
 {REFERENCES}{REFERENCES}\list
 {[\arabic{enumi}]}{\settowidth\labelwidth{[#1]}\leftmargin\labelwidth
 \advance\leftmargin\labelsep
 \usecounter{enumi}}
 \def\newblock{\hskip .11em plus .33em minus -.07em}
 \sloppy
 \sfcode`\.=1000\relax}
\def\upcite#1{\raise6pt\hbox{\scriptsize
\cite{#1}}}
  \def\lsim{\mathrel {\vcenter {\baselineskip 0pt \kern 0pt
    \hbox{$<$} \kern 0pt \hbox{$\sim$} }}}
    \def\gsim{\mathrel {\vcenter {\baselineskip 0pt \kern 0pt
    \hbox{$>$} \kern 0pt \hbox{$\sim$} }}}

\def\hline{\noalign{\hrule \vskip2pt}}

\def\|{\ifmmode\Vert\else \char`\|\fi}
\ifx\oldzeta\undefined                          
  \let\oldzeta=\zeta                            
  \def\zzeta{{\raise 2pt\hbox{$\oldzeta$}}}     
  \let\zeta=\zzeta                              
\fi

\ifx\oldchi\undefined                           
  \let\oldchi=\chi                              
  \def\cchi{{\raise 2pt\hbox{$\oldchi$}}}       
  \let\chi=\cchi                                
\fi

\def\frac#1#2{{#1 \over #2}}

\def\half{\ifinner {\scriptstyle {1 \over 2}}
   \else {1 \over 2} \fi}

\def\simge{\mathrel{%
   \rlap{\raise 0.511ex \hbox{$>$}}{\lower 0.511ex \hbox{$\sim$}}}}
\def\simle{\mathrel{
   \rlap{\raise 0.511ex \hbox{$<$}}{\lower 0.511ex \hbox{$\sim$}}}}

\def\buildchar#1#2#3{{\null\!                   
   \mathop#1\limits^{#2}_{#3}                   
   \!\null}}                                    
\def\overcirc#1{\buildchar{#1}{\circ}{}}

\def\slashchar#1{\setbox0=\hbox{$#1$}           
   \dimen0=\wd0                                 
   \setbox1=\hbox{/} \dimen1=\wd1               
   \ifdim\dimen0>\dimen1                        
      \rlap{\hbox to \dimen0{\hfil/\hfil}}      
      #1                                        
   \else                                        
      \rlap{\hbox to \dimen1{\hfil$#1$\hfil}}   
      /                                         
   \fi}                                         %

\def\subrightarrow#1{
  \setbox0=\hbox{
    $\displaystyle\mathop{}
    \limits_{#1}$}
  \dimen0=\wd0
  \advance \dimen0 by .5em
  \mathrel{
    \mathop{\hbox to \dimen0{\rightarrowfill}}
       \limits_{#1}}}                           

\def\overlay#1#2{\ifmmode%
\setbox0=\hbox{$#1$}%
\setbox1=\hbox to\wd0{\hss$#2$\hss}\else%
\setbox0=\hbox{#1}%
\setbox1=\hbox to\wd0{\hss#2\hss}\fi%
#1\hskip-\wd0\box1 }

\def\pmb#1{\leavevmode\setbox0=\hbox{#1}%
\kern-.02em\copy0\kern-\wd0
\kern.04em\copy0\kern-\wd0
\kern-.02em\raise.04em\box0 }

\def\vereq#1#2{\lower3pt\vbox{\baselineskip1.5pt \lineskip1.5pt
\ialign{$\m@th#1\hfill##\hfil$\crcr#2\crcr\sim\crcr}}}

\def\tensor#1{\protect\@ontopof{#1}{\leftrightarrow}{1.15}\mathord{\box2}}
\def\overstar#1{\protect\@ontopof{#1}{\ast}{1.15}\mathord{\box2}}
\def\overdots#1{\protect\@ontopof{#1}{\cdots}{1.0}\mathord{\box2}}
\def\overcirc#1{\protect\@ontopof{#1}{\circ}{1.2}\mathord{\box2}}
\def\loarrow#1{\protect\@ontopof{#1}{\leftarrow}{1.15}\mathord{\box2}}
\def\roarrow#1{\protect\@ontopof{#1}{\rightarrow}{1.15}\mathord{\box2}}

\def\@ontopof#1#2#3{%
{\mathchoice
{\@@ontopof{#1}{#2}{#3}\displaystyle\scriptstyle}%
{\@@ontopof{#1}{#2}{#3}\textstyle\scriptstyle}%
{\@@ontopof{#1}{#2}{#3}\scriptstyle\scriptscriptstyle}%
{\@@ontopof{#1}{#2}{#3}\scriptscriptstyle\scriptscriptstyle}%
}%
}

\def\@@ontopof#1#2#3#4#5{%
\setbox0=\hbox{$#4#1$}%
\setbox1=\hbox{$#5#2$}%
\setbox2=\hbox{}\ht2=\ht0 \dp2=\dp0 %
\ifdim\wd0>\wd1 %
\setbox1=\hbox to\wd0{\hss\box1\hss}%
\mathord{\rlap{\raise#3\ht0\box1}\box0}%
\else   %
\setbox1=\hbox to.9\wd1{\hss\box1\hss}%
\setbox0=\hbox to\wd1{\hss$#4\relax#1$\hss}%
\mathord{\rlap{\copy0}\raise#3\ht0\box1}%
\fi
}%

\def\lambdabar{\protect\@lambdabar}
\def\@lambdabar{%
\relax
\bgroup
\def\@tempa{\hbox{\raise.73\ht0
\hbox to0pt{\kern.25\wd0\vrule width.5\wd0
height.1pt depth.1pt\hss}\box0}}%
\mathchoice{\setbox0\hbox{$\displaystyle\lambda$}\@tempa}%
{\setbox0\hbox{$\textstyle\lambda$}\@tempa}%
{\setbox0\hbox{$\scriptstyle\lambda$}\@tempa}%
{\setbox0\hbox{$\scriptscriptstyle\lambda$}\@tempa}%
\egroup
}

\def\corresponds{{\lower.2ex\hbox{=}}{\rm\kern-.75em^\triangle}}
\def\succsim{\succ\kern-.9em_\sim\kern.3em}
\def\precsim{\prec\kern-1em_\sim\kern.3em}
\def\slantfrac#1#2{\kern1em^{#1}\kern-.3em/\kern-.1em_{#2}}

\begin{document}

\begin{center}
{\Large\bf Diffraction as a Consequence of Faraday's Law}
\\

\medskip

Max. S.~Zolotorev
\\
{\sl Center for Beam Physics, Lawrence Berkeley National Laboratory,
Berkeley, CA 94720}
\\
Kirk T.~McDonald
\\
{\sl Joseph Henry Laboratories, Princeton University, Princeton, NJ 08544}
\\
(Jan.\ 11, 1999)
\end{center}

\section{Problem}

A linearly polarized plane electromagnetic wave of frequency $\omega$
is normally incident on an opaque screen with a square aperture of edge
$a$.

Show that the wave has a longitudinal magnetic field once it has passed
through the aperture by an application of Faraday's Law to a loop
parallel to the screen, on the side away from the source.  Deduce the
ratio of longitudinal to transverse magnetic field, which is a 
measure of the diffraction angle.

\section{Solution}

Consider a linearly polarized wave with electric field 
${\bf E}_x e^{i(kz - \omega t)}$ incident on
a perfectly absorbing screen in the plane $z = 0$ with a square aperture of
edge $a$ centered on the origin. 
We apply the integral form of Faraday's Law
to a semicircular loop with its straight edge bisecting the aperture
and parallel to 
the transverse electric field ${\bf E}_x$, as shown in the figure.
The electric field is essentially zero
close to the screen on the side away from the source.
Then, at time $t = 0$,
\begin{equation}
\oint {\bf E} \cdot d{\bf l} \approx E_x \, a \neq 0.
\label{eq1}
\end{equation}
If the loop were on the source side of the screen, the integral would
vanish. 

\begin{figure}[htp]  
\begin{center}
\includegraphics[width=3in, angle=0, clip]{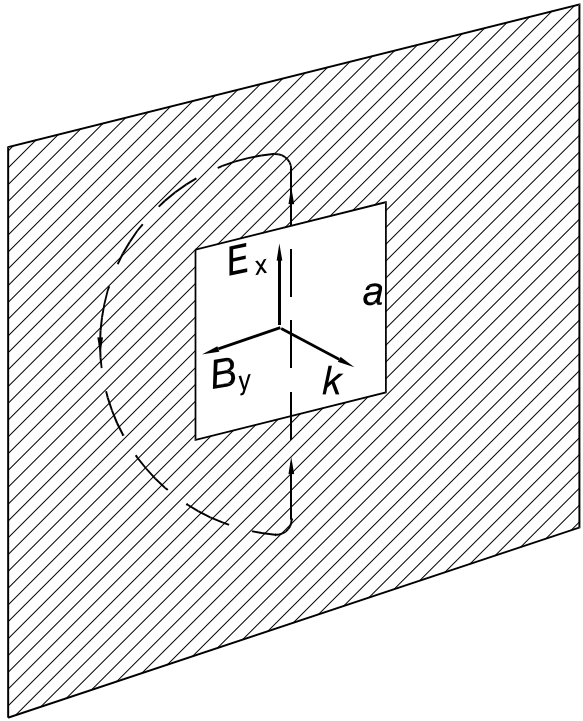}
\parbox{5.5in} 
{\caption[ Short caption for table of contents ]
{\label{fig1} A screen with a square aperture of edge $a$ is illuminated by
a linearly polarized electromagnetic wave.  
The imaginary loop shown by the dashed curve lies close to the screen, on
the side away from the source, and so is
partly in the shadow of the wave.
}}
\end{center}
\end{figure}

Faraday's Law tells us immediately that the time derivative of the
magnetic flux through the loop
is nonzero.  Hence, there must be a nonzero longitudinal component,
$B_z$,
to the magnetic field, once the wave has passed through the aperture.

In Gaussian units,
\begin{equation}
B_y \, a = E_x \, a \approx \oint {\bf E} \cdot d{\bf l}
= - {1 \over c} {d \over dt} \int {\bf B} \cdot d{\bf S}
\approx - {1 \over c} {dB_z \over dt} {a^2 \over 2},
\label{eq2}
\end{equation}
where $B_z$ is a characteristic value of the longitudinal component of
the magnetic field over that half of the aperture enclosed by the loop.
The longitudinal magnetic field certainly has time dependence of the form
$e^{-i \omega t}$, so $dB_z/dt = -i\omega B_z = -2 \pi i c B_z / 
\lambda$, and eq.~(\ref{eq2}) leads to 
\begin{equation}
{B_z \over B_y} \approx -  {i \lambda \over \pi a}.
\label{eq3}
\end{equation}
By a similar argument for a loop that enclosed the other half of the aperture,
$B_z / B_y \approx i \lambda/ \pi a$ in that region; $B_z = 0$ in the
plane $y = 0$.

We see that the wave is no longer a plane wave after passing through
the aperture, and we can say that it has been diffracted as a consequence of
Faraday's Law.

This argument emphasizes the fields near the aperture.  A detailed
understanding of the fields far from the aperture requires more than just
Faraday's Law.  A simplified analysis is that that magnitude of the
ratio (\ref{eq3}) is a measure of the spread of angles of the magnetic field
vector caused by the diffraction, and so in the far zone the wave occupies a
cone of characteristic angle $\lambda/ \pi a$.

\section{Comments}

Using the fourth Maxwell equation
including the displacement current, we can make an argument
for diffraction of the electric field similar to that given above for the
magnetic field.

After the wave has passed through the aperture of size $a$, it is very
much like 
a wave that has been brought to a focus of size $a$.  Hence, we learn that
near the focus $(x,y,z) = (0,0,0)$ of a linearly polarized electromagnetic 
wave with ${\bf E} = E \hat{\bf x}$ and propagating in the $z$ direction, 
there are both longitudinal electric and magnetic fields, and that 
$E_z$ and $B_z$ are antisymmetric about the planes $x = 0$ and $y = 0$, 
respectively.

Also, eq.~(\ref{eq3}) indicates that near the focus the longitudinal and
transverse fields are $90^\circ$ out of phase. 
Yet, far from the focus, the transverse and longitudinal fields become in phase,
resulting in spherical wavefronts that extend over a cone of characteristic
angle $\lambda/ \pi a$.  For this to hold, the longitudinal and the
transverse fields must experience phase shifts that differ by $90^\circ$ 
between the focal point and the far zone.

It is only a slight leap from the present argument to conclude that the 
transverse fields undergo the extra phase shift.  This was first deduced 
(or noticed) by Guoy \cite{Siegman} in 1890 via the 
Huygens-Kirchhoff integral \cite{Landau}.  The latter tells us that
the secondary wavelet $\psi$ at a large distance $r$ from a small region 
of area $A$ where the wave amplitude is $\psi_0 e^{-i \omega t}$ is
\begin{equation}
\psi = {k  \psi_0 A \over 2 \pi i} {e^{i(kr - \omega t)} \over r} 
= {k \psi_0 A \over 2 \pi } {e^{i(kr - \omega t - \pi/2)} \over r}.
\label{eq5}
\end{equation}
The possibly mysterious factor of $i$ in the denominator of the 
Huygens-Kirchhoff integral
implies a $90^\circ$ phase shift between a focus and the far field of a beam of
light.  Here, we have seen that this phase shift can also be considered as
a consequence of Faraday's Law.

\end{document}